# Developing a Maximum-Entropy Restricted Boltzmann Machine with a Quantum Thermodynamics Formalism

Roshawn Terrell[1], Eleanor Watson[2], Timofey Golubev[3]

[1]Vilnius University, Lithuania
[2]Atlantic Council GeoTech Center, UK
[3]Michigan State University, USA

Email: [1]roshawn.terrell@ff.stud.vu.lt, [2]nwatson@atlanticcouncil.org, [3]golubevt@msu.edu

## Abstract

We propose a Restricted Boltzmann Machine (RBM) neural network using a quantum thermodynamics formalism and the maximization of entropy as the cost function for the optimization problem. We verify the possibility of using an entropy maximization approach for modeling physical systems by applying it to the Inverse Ising Problem, which describes a system of interacting spins. The temperature-dependent behaviors of physical quantities, such as magnetization and Gibbs energy, were calculated from the Ising Hamiltonian whose coefficients were obtained through entropy maximization and were found to be in good agreement to literature results. These results suggest that RBM neural networks using the principle of maximum entropy can be applied to modeling physical systems that can be described by discrete states, including fundamental quantum physics such as topological systems, and biological systems such as the correlated spiking of neurons.

## 1. Introduction

There has been a long history of physicists exploring analogies between statistical mechanics and the dynamics of neural networks. For example, the results of a convolutional autoencoder applied to the 3D Ising model predicted the magnetization and susceptibility as functions of temperature with high accuracy when compared with results from other numerical methods for the Ising model [1]. In another work, a feedforward neural network was used to calculate the ground state of the Bose-

Hubbard model and the results were in good agreement to other successful numerical methods, such as exact diagonalization and the Gutzwiller approximation [2]. It also has been shown that a causal generalization of entropic forces suggests the possibility to create a general thermodynamic model of adaptive behavior as a nonequilibrium process in open systems [3]. In the context of biological systems, entropy-based formalisms have been used to describe the time evolution of the state of the system based on asynchronous parallel processing [4]. Additionally, Maximum Entropy models have been used to describe biological systems, such as describing how local interactions can generate collective dynamics in neural activity [5, 6, 7] and describing amino acid interactions in proteins [8]. It has been shown that the analogy between statistical mechanics and the dynamics of neural networks can be turned into a precise mapping and connected to experimental data. For example, the pairwise interactions Ising model provides a very accurate description of the patterns of neural firing (spiking and silence) in the retina, even though it excludes all higher order interactions between multiple cells [9, 6, 7]. The Ising model can be considered as a simplified Restricted Boltzmann Machine, which is a two-layer neural network that makes up the building blocks of deep-belief networks for unsupervised machine learning. In this work, motivated by the aforementioned works, we propose a maximum-entropy model for Restricted Boltzmann Machine neural networks using an Ising-type Hamiltonian in the quantum thermodynamics formalism.

## 2. Maximal Entropy Bose-Hubbard RBM Model for Neurons

### 2.1 Model Formalism

A large system of interacting particles can be described by a Hamiltonian of the form

$$H_{tot} = H_{sys}(x, \lambda(t)) + H_{bath}(y) + h_{int}(x, y) \qquad (1)$$

where $x$ accounts for all the coordinate degrees of freedom for the particles in the system, y does the same for the bath, and the Hamiltonian functions $H_{sys}$, $H_{bath}$, and $h_{int}$ define conservative interactions among the various position coordinates of the system and the bath [2, 10, 11]. The function $\lambda(t)$ is a time-varying external field that acts exclusively on the system and can do work on the coordinates $x$. The interaction term, $h_{int}$, is generally assumed to be small and ignored.

In this work, we utilize the Bose-Hubbard model, which provides the simplest description of interacting spinless bosons (particles that obey Bose-Einstein statistics) on a lattice and is widely used to describe the phase transition to the state of Bose-Einstein condensate in systems of ultracold bosons in an optical lattice [11, 12, 13]. The physics of the model is given by the Bose-Hubbard Hamiltonian

$$\hat{H} = -J \sum_{\langle i,j \rangle} \hat{b}_i^\dagger \hat{b}_j + \frac{U}{2} \sum_i \hat{n}_i (\hat{n}_i - 1) - \mu \sum_i \hat{n}_i \qquad (2)$$

where $\langle i,j \rangle$ denotes summation over all neighboring lattice sites, $\hat{b}_i^\dagger$ is the creation operator for bosons on the state $i$, $\hat{b}_j$ is the annihilation operator for bosons on state $j$, and $\hat{n}_i = \hat{b}_i^\dagger \hat{b}_i$ is the number operator, which counts the number of bosons on site $i$. The model is parametrized by the nearest-neighbor interaction strength $J$, the on-site interaction strength $U$, and the chemical potential $\mu$, which sets the total number of particles in the system. The first term describes the kinetic energy of particles due to hopping where $\hat{b}_i^\dagger \hat{b}_j$ describes the annihilation of a boson on lattice site $j$ and creation of a boson on site $i$; in other words, the boson "hops" from site $j$ to $i$. The second term describes the interaction energy between bosons on the same site (this energy is zero when there is no or only one boson). For $n_i$ particles on site $i$, the number of interacting pairs is $\frac{1}{2} n_i (n_i - 1)$ and each pair contributes $U$ to the energy.

Similar models have been used to describe the behavior of the system in complex network topologies [14] and for networks in machine learning [15]. In these previous works, the authors used the Mean-Field Approximation (MFA) to simplify the formalism, while we propose to use the Hubbard operator (X-operator) approach [16], which will allow us to take into account the interaction between bosons in a more reliable way than possible with the MFA. The simplest representation of the Hubbard operators can be written as the outer product of basis vectors. In Dirac notation

$$\hat{X}_i^{nm} = |n,i\rangle\langle m,i| \qquad (3)$$

Then, the creation and annihilation operators can be written in terms of X-operators as

$$\hat{b}_i = \sum_n \sqrt{n+1}\, \hat{X}_i^{n,n+1}, \quad \hat{b}_i^\dagger = \sum_n \sqrt{n+1}\, \hat{X}_i^{n+1,n} \qquad (4)$$

Thus, the Bose-Hubbard Hamiltonian (Equation 2) can be reformulated as

$$\hat{H} = -J \sum_{\langle i,j \rangle} \left( \sum_n \sqrt{n+1}\, \hat{X}_i^{n+1,n} \right) \left( \sum_n \sqrt{n+1}\, \hat{X}_j^{n,n+1} \right) + \sum_{i,n} \alpha_n \hat{X}_i^{nn} \qquad (5)$$

where $n$ is the number of neurons on each site $i$ and

$$\alpha_n = \frac{U}{2} n(n-1) - \mu n, \qquad (6)$$

are single-site energies. One can write Equation 5 in matrix form as $H = \sum_i H_i^{latt}$ where

$$\hat{H}_{latt,i} = \begin{pmatrix} \alpha_0 & -J & 0 & 0 & ... \\ -J & \alpha_1 & -\sqrt{2}J & 0 & ... \\ ... & -\sqrt{2}J & \alpha_2 & -\sqrt{3}J & ... \\ ... & ... & ... & ... & ... \\ .... & .... & ... & -\sqrt{n}J & \alpha_n \end{pmatrix} \qquad (7)$$

The possible energies associated with each site are given by the eigenvalues of the Hamiltonian, in other words, the values $\beta_k$ that satisfy $\hat{H}_{latt,i}|\psi> = \beta_k |\psi>$. This is a symmetric tridiagonal matrix and can be diagonalized using standard numerical methods. Diagonalization will result in a matrix of the form

$$\hat{H}'_{latt,i} = \begin{pmatrix} \beta_0 & 0 & 0 & 0 \\ 0 & \beta_1 & 0 & 0 \\ 0 & 0 & ... & 0 \\ 0 & 0 & 0 & \beta_n \end{pmatrix} \qquad (8)$$

where the new functions $\beta_k$ are functions of $\alpha_k$ and $J$ and can be interpreted as energy levels.

In statistical mechanics, the possible states of a system of particles in thermodynamic equilibrium with a heat bath are described by the grand canonical ensemble. The grand canonical ensemble assigns a probability $P_k = \exp\left[\frac{\Omega + \mu N - \beta_k}{k_b T}\right]$ to each distinct state of the system, where $\Omega = \Omega(\mu, V, T)$ is the grand canonical potential (GCP), N is the number of particles, $k_b$ is Boltzmann's constant, and T is temperature [17]. Many important ensemble average quantities, including entropy, can be calculated from the derivatives of the GCP, which is defined as

$$\Omega = k_b T ln\left(\sum_k \exp\left(-\frac{\beta_k}{k_b T}\right)\right) \qquad (9)$$

Entropy can be calculated from the derivative of the GCP with respect the temperature ($T$) as follows

$$S = -\frac{\partial \Omega}{\partial T} = -k_b ln\left(\sum_k \exp\left(-\frac{\beta_k}{k_b T}\right)\right) \\ + k_b T \frac{1}{\left(\sum_k \exp\left(-\frac{\beta_k}{k_b T}\right)\right)} \left(\sum_k \exp\left(-\frac{\beta_k}{k_b T}\right)\left(\frac{\beta_k}{k_b T^2}\right)\right) \qquad (10)$$

Equation 10 can be solved numerically using the functional relationship

$$\beta_k = f(\alpha_{k-1}, \alpha_k, \alpha_{k+1}, J) \qquad (11)$$

Here, we have the dependence only from α with indexes $k-1, k, k+1$ because in general, the resulting matrix (Equation 7) is tridiagonal.

We limit the remainder of our discussion to the known physics of Bose-Einstein systems, the so-called hard-core bosons approximation. For the case of the Bose-Hubbard model this means that the only possible states are those where the number of bosons on each site is either 0 or 1. Thus, there are only two possible energy eigenvalues for each site, reducing the Hamiltonian from Equation 7 to a 2x2 matrix

$$\hat{H}_{latt,i} = \begin{pmatrix} \alpha_0 & -J \\ -J & \alpha_1 \end{pmatrix} = \begin{pmatrix} 0 & -J \\ -J & -\mu \end{pmatrix} \tag{12}$$

where we have substituted for the values of $\alpha_0$ and $\alpha_1$ using Equation 7. The energy eigenvalues ($\beta_k$) of this Hamiltonian are

$$\beta_0 = -\frac{\mu}{2} - \sqrt{\frac{\mu^2}{4} + J^2} \text{ and } \beta_1 = -\frac{\mu}{2} + \sqrt{\frac{\mu^2}{4} + J^2} \tag{13}$$

Then, the GCP can be written as

$$\Omega = k_b T \ln\left(\exp\left(-\frac{\beta_0}{k_b T}\right) + \exp\left(-\frac{\beta_1}{k_b T}\right)\right) \tag{14}$$

and the entropy is

$$S = -\frac{\partial \Omega}{\partial T} = -\left[k_b \ln\left(\exp\left(-\frac{\beta_0}{k_b T}\right) + \exp\left(-\frac{\beta_0}{k_b T}\right)\right) + \frac{1}{T}\frac{\left(\exp\left(-\frac{\beta_0}{k_b T}\right)\beta_0 + \exp\left(-\frac{\beta_1}{k_b T}\right)\beta_1\right)}{\exp\left(-\frac{\beta_0}{k_b T}\right) + \exp\left(-\frac{\beta_0}{k_b T}\right)}\right] \tag{15}$$

Thus, we have expressed the entropy as a function of the model parameters

$$S = \Phi(T, \mu, J) \tag{16}$$

For maximizing entropy, we can set the derivatives of this function with respect to the model parameters to be equal to zero (for example, $\frac{\partial \Phi}{\partial T} = 0$ if we want find critical temperature) to find the behavior of the system with respect to criticality. Figure 1 presents the dependence of entropy on the tunneling coefficient ($J$) for different values of the chemical potential ($\mu$) under the hard-core bosons approximation. All variables are in units of energy (for this we use $k_B T \to 1$). As one can see from Figure 1, the entropy has a maximum at the value $J = 0$ for any value of the chemical potential. In cases beyond the hard-core bosons approximation, where the number of bosons on a site can be more than one, we will have one more important parameter $U$, which describes the on-site interaction of the bosons, and the behavior of the entropy function will be more complex.

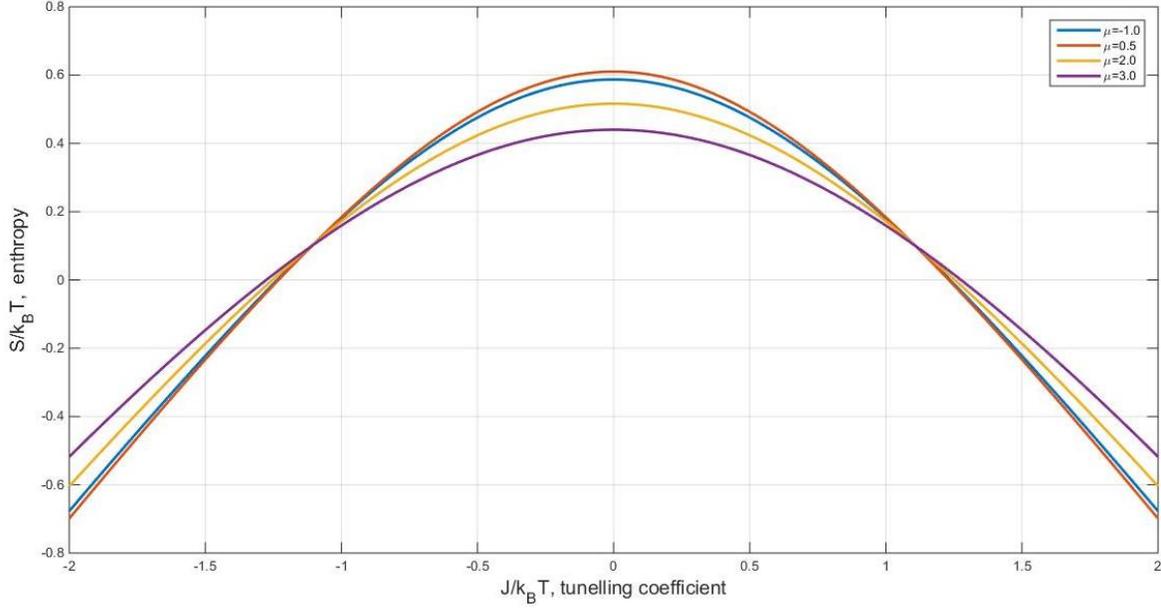

**Figure 1. Entropy the Bose-Hubbard model under the hard-core bosons approximation as a function of tunneling coefficient *J*, for different values of chemical potential *μ*.**

**2.2 Constructing the RBM Neural Network**

Boltzmann Machines are bidirectionally connected neural networks of stochastic processing units that can be used to learn important aspects of an unknown probability distribution based on data samples from the distribution [18]. Note that unlike other commonly-used deterministic neural networks (e.g. multilayer perceptron), which estimate a value based on inputs, Boltzmann Machines estimate the probability densities of the inputs. Boltzmann Machines are shallow networks with only two layers: a visible layer and a hidden layer. The units in the visible layer (visible units) correspond to the components of an observation and the units in the hidden layer (hidden units) model dependencies between these components. While the learning process is impractical in general Boltzmann Machines due to high computational expense, it can be made quite efficient by defining a Restricted Boltzmann Machine (RBM) that does not allow connections between units of the same type (i.e. hidden-hidden and visible-visible connections are not allowed). An RBM is mathematically defined by a Hamiltonian that has the form

$$\widehat{H}_{RBM} = \sum_{l} a_l v_l + \sum_{k} b_k h_k + \sum_{k,l} W_{kl} v_l h_k \tag{17}$$

where $v_l$ denote the visible units, $h_k$ denote the hidden units, and $a_l$, $b_k$, and $W_{kl}$ are the weights that define the neural network's architecture.

The numerical construction of our artificial neural network (ANN) is based on the work of Bausch and Leditzky, who used neural networks to represent Absolutely Maximally Entangled (AME) quantum states [19]. Bausch and Leditzky formulated an optimization problem for AME states for dimension $d$ and number of qubits $n$ as follows. The AME quantum state can be defined by $|\psi_{n,d}\rangle$. This state can be expressed as a decomposition with respect to a known basis set $\{|i\rangle\}_{i=0}^{d-1}$

$$|\psi_{n,d}\rangle = \frac{1}{C}\sum \psi(i^n)|i^n\rangle \tag{18}$$

where $C$ is a normalization constant and the function $\psi(i^n)$ is computed by the neural network. There are several different approaches that can be used to encode the strings $i^n$ (e.g., binary, scaled, or one-hot). These basis strings $\{|i\rangle\}_{i=0}^{d-1}$ are the visible and hidden units from the RBM and also known as the "spin-list", which is some space of variables that are encoded to be used in the neural network. For example, with binary encoding for an AME quantum state with (n=3, d=2) we will get the following spin list:

```
0  0  0  0  1  1  1  1
0  0  1  1  0  0  1  1
0  1  0  1  0  1  0  1
```

Here we have $d^n$ columns of binary encoding of basis strings $\{|0\rangle\}$, $\{|1\rangle\}$, and $\{|2\rangle\}$.

The RBM Hamiltonian has the same form as our Bose-Hubbard Hamiltonian from Equation 2

$$\hat{H} = -J\sum_{<i,j>} \hat{b}_i^\dagger \hat{b}_j + \frac{U}{2}\sum_i \hat{n}_i(\hat{n}_i - 1) - \mu\sum_i \hat{n}_i \tag{19}$$

therefore, allowing the extension of the ANN developed by Bausch and Leditzky to our model. The wavefunctions for the RBM can be written in terms of binary AME states as

$$|\psi_n\rangle = \sum_{i^n \in [0,1]^n} \psi_n(i^n)|i^n\rangle \tag{20}$$

where $|i^n\rangle = |i_1\rangle \otimes |i_2\rangle \otimes |i_3\rangle \otimes \ldots$ is the spin list. Then, the wave functions for the RBM can be written as

$$|\psi_{RBM}\rangle = |\psi_n\rangle = \sum_{i^n \in [0,1]^n} \sum_{h^n \in \{0,1\}} \frac{\exp(-E(i^n, h^n))}{Z} |i^n\rangle \tag{21}$$

where Z is the partition function and equal to the sum of all possible combinations

$$Z = \sum_{i,h} \exp(-E(i^n, h^n)) \tag{22}$$

We can formulate the optimization problem as follows. For a subset $S$ of $n$ states, the constraint on $|\psi_n\rangle$ to be being absolutely maximally entangled is related to the average linear entropy

$$Q_m(\psi_n)) = \frac{1}{n} \sum_S \frac{2^n}{2^n - 1} (1 - tr(\rho_S^2)) \tag{23}$$

where $\rho_S = tr(|\psi_n\rangle)$. Equation 23 is the main equation for optimization. Optimization approaches such as artificial bee colony, pattern search, and gradient search can be used to optimize $Q_m(\psi_{n,d})$ and generate the ANN parameters and state functions ($\psi_n$) that obey the maximum entropy condition.

### 2.3. Simplification to the Ising Model

The RBM can be simplified to an Ising model Hamiltonian, borrowed from condensed matter physics. If we assume that the visible and hidden units are equivalent (let's denote them as $v_l = h_k \equiv s_i$), then, taking the opposite sign to the weights, we can rewrite Equation 2 as

$$\hat{H} = -\sum_{i=1}^N a_i s_i - \sum_{i,j} W_{i,j} s_i s_j \tag{24}$$

This has the form of the Hamiltonian of the Ising Model when we interpret the visible and hidden units as spins $s_{i/j}$. We can rewrite the Ising Model Hamiltonian using the more common notation from condensed matter physics [20]

$$\hat{H} = -\sum_{i,j} J_{ij} \sigma_i \sigma_j - \sum_i h_i \sigma_i \tag{25}$$

where $\sigma_i$ is the spin with possible orientations (up is $\sigma_i = 1$ and down is $\sigma_i = -1$), $h_i$ is the external magnetic field, and $J_{ij}$ is the pairwise coupling between spins. Figure 2 shows a depiction of this form of the Ising model with different values of magnetic field on every site, as well as possible pair correlations between all sites. It can also be described as a 1D spin chain with periodic boundaries and fully pairwise interactions. In terms of biological neurons, the spins can be re-interpreted as the "on" or "off" state of neurons where the "on" state is when the neuron is emitting an action potential (electrochemical pulse).

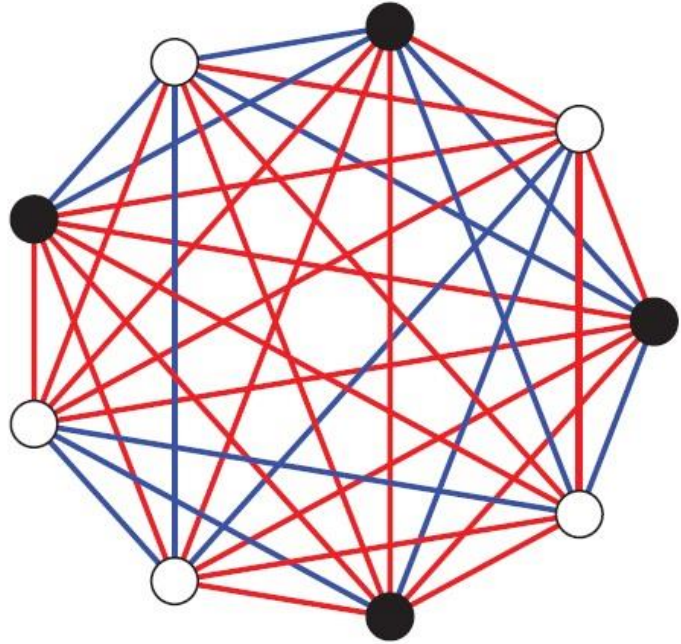

**Figure 2: Fully connected pairwise Ising model. Each spin can have a value equal to -1 (black) or 1 (white), and interact with every other spin with an attractive (red) or repulsive (blue) value. From** [20]**.**

The Restricted Boltzmann Machine uses entropy maximization to the find the optimal set of coefficients $a_l, b_k, W_{kl}$ for its Hamiltonian (Equation 17). Similarly, the Inverse Ising problem uses entropy maximization to find the find the optimal set of magnetic field $h_i$ and interaction $J_{ij}$ coefficients for its Hamiltonian (Equation 25). Furthermore, the RBM and Ising model are identical when we interpret the possible values of hidden/visible units in the RBM as values of spins in the Ising model. Therefore, as a first step, our proposed entropy-based approach to describe a system of neurons can be verified by applying it to the simpler Ising model.

## 3. Results and Discussion

For preliminary verification of an entropy-based approach of describing a system of neurons, we solved the Inverse Ising Problem; in other words, we obtain coefficients of the Ising Hamiltonian through optimization under the maximum entropy condition. The numerical calculations were performed using the coniii-3 package [20]. We used N=4, 6, and 8 for the possible numbers of sites in the Ising Model. This choice can be explained by the fact that even numbers are more suitable from symmetry considerations and N=2 is too small for good reliability of an inverse solution. The average values of the Ising model parameters over all spins (average field $\langle h \rangle$ and average interaction $\langle J \rangle$), which are found by solving the Inverse Ising problem, are shown in Table 1.

**Table 1. Calculated values of average field $\langle h \rangle$ and average pairwise interaction <J> for Ising model with N = 4, 6, and 8 sites.**

| N | Average Magnetization $\langle h \rangle$ | Average Pairwise interaction $\langle J \rangle$ |
|---|---|---|
| 4 | 0.126744 | 0.044978 |
| 6 | 0.106630 | 0.013432 |
| 8 | 0.085627 | 0.013398 |

These Ising Hamiltonian coefficients can be interpreted as thermodynamics quantities. In the Ising model, given known values of $h_i$ and $J_{ij}$, we can express the system energy term for every possible combination of spins as

$$E(\sigma) = -\sum_{i,j} J_{ij}\sigma_i\sigma_j - \sum_i h_i\sigma_i \tag{26}$$

If we want to calculate the average of some observable variable, we have to apply statistical averaging using

$$\langle A \rangle = \frac{\sum_\sigma A \exp\left(-\frac{E(\sigma)}{T}\right)}{Z_N} \tag{27}$$

where $Z_N$ is the partition function for the system of N sites and is a sum over all possible site configurations (for $N$ sites we have $2^N$ configurations)

$$Z_N = \sum_\sigma \exp\left(-\frac{E(\sigma)}{T}\right) \tag{28}$$

where we use the 'energy' form of temperature which is $T \equiv k_B T$. For example, the magnetization per site is calculated by

$$\langle m \rangle = \frac{1}{N} \sum_\sigma \frac{exp\left(-\frac{E(\sigma)}{T}\right)}{Z_N} \sum_i \sigma_i \qquad (29)$$

Furthermore, another fundamental thermodynamics quantity, the Gibbs entropy ($S$) of the system, can be derived from the free energy ($F$) and partition function ($Z_N$) as follows

$$S = -\frac{\partial F}{\partial T} = \ln(Z_N) + \frac{T}{Z_N}\frac{\partial Z_N}{\partial T} = \ln(Z_N) - \frac{1}{Z_N T} \sum_\sigma E(\sigma) \exp\left(-\frac{E(\sigma)}{T}\right) \qquad (30)$$

where $F$ is defined as $F = -T\ln(Z_N)$.

We calculated the values for average magnetization, free energy, and Gibbs entropy as functions of temperature T using the values of magnetic field (H) and pairwise interaction (J) obtained from the inverse Ising problem. The value of temperature was taken in units of maximal value of the inter-site interaction, which is the standard procedure for Ising-like models. The results (Figure 3) are in good agreement with known results from literature [11, 4]. More specifically, our results for the behavior of average magnetization (Fig. 3a) are consistent with those from [11], where a 2D Ising model with local distortions was investigated analytically. Our results for the temperature dependence of entropy are similar to that from [4], where numerical linked-cluster (NLC) algorithms were used for the calculation.

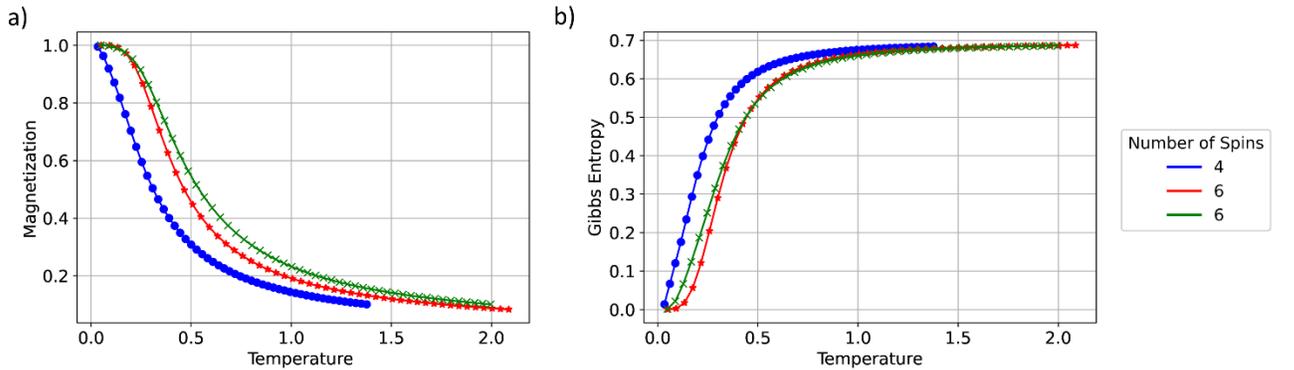

**Figure 3: Fundamental thermodynamic quantities calculated from the solution to the Inverse Ising Problem for a system of 4, 6, and 8 spins. a) Average magnetization per site and b) Gibbs entropy as functions of temperature.**

## Conclusions

Using maximum entropy as the optimization function, one can construct the Restricted Boltzmann Machine that describes the physical system that is being modeled. Solving this optimization problem will provide the optimal parameters for different possible physical representations of the system. In this work, we described the theoretical formalism of a Bose-Hubbard RBM model for neural networks. We verified the concept of a maximum entropy approach by testing it on the simpler Inverse Ising Problem where we compared results for fundamental thermodynamical quantities from the Ising Hamiltonian description of a neural network (a simplified RBM) to results from the literature. In future work, the proposed maximum entropy Bose-Hubbard model for neural networks should be implemented and tested. The model is expected to be applicable for investigations into fundamental quantum physics, biological systems, and could be generalized for modeling other complex systems, such as financial markets.